\documentclass[%
 amsmath,amssymb,
 aps,
]{revtex4-1}

\usepackage{graphicx}
\usepackage{dcolumn}
\usepackage{bm}
\usepackage{makecell}
\usepackage{rotating}
\usepackage{float}
\usepackage{longtable,verbatim}
\usepackage{tabularx}
\usepackage{multirow}

\newcommand{\reqn}[1]{Eq.~\ref{eq:#1}}
\renewcommand{\v}[1]{{{\bf #1}}}

\usepackage[normalem]{ulem}
\usepackage{soul}
\usepackage{xcolor}

\begin{document}

\preprint{APS/123-QED}

\title{Statistical estimation theory detection limits for label-free imaging}

\author{Lang Wang}%
\author{Maxine Xiu}%
\author{Ali Pezeshki}%
\author{Randy Bartels}
 \email{rbartels@morgridge.org}
\affiliation{Colorado State University, Fort Collins, CO 80523}%

\date{\today}

\begin{abstract}
The emergence of label-free microscopy techniques has significantly improved our ability to precisely characterize biochemical targets, enabling non-invasive visualization of cellular organelles and tissue organization. Each label-free method has specific benefits, drawbacks, and varied varied sensitivity under measurement conditions across different types of specimens. To link all these disparate label-free optical interactions together and to compare detection sensitivity of these modalities, we investigate their sensitivity within the framework of statistical estimation theory. This paper introduces a comprehensive unified framework for evaluating the bounds for signal detection with label-free microscopy methods, including second harmonic generation (SHG), third harmonic generation (THG), coherent anti-Stokes Raman scattering (CARS), coherent Stokes Raman scattering (CSRS), stimulated Raman loss (SRL), stimulated Raman gain (SRG), stimulated emission (SE), impulsive stimulated Raman scattering (ISRS), transient absorption (TA), and photothermal effect (PTE). A general model for signal generation induced by optical scattering is developed. Based on this model, the information obtained is quantitatively analyzed using Fisher information, and the fundamental constraints on estimation precision are evaluated through the Cram\'er-Rao Lower Bound (CRLB).
\end{abstract}

\maketitle


\section{\label{sec:intro} Introduction}


Optical imaging provides a method of observing biological systems that is particularly powerful for studying dynamics in live specimens. Information obtained from optical microscopes is derived from the light collected from the specimen. In one widespread approach, exogenous labels (often molecular dyes or fluorophores) are applied to interrogate the behavior of cells and tissues, such as nucleic acids, cytoplasm, extracellular proteins, or particular biomolecules. Despite the incredible power that comes from the specificity of the application of these labels, the labels carry their own problems. Many are toxic or severely disrupt biological function, complicating interpretation of data from imaging experiments. In addition, the introduction of external labels is often impeded by physical processes, such as the need for labels to diffuse through tissue or to pass through the blood-brain barrier.

An alternative strategy for optical microscopy, label-free imaging, uses intrinsic optical properties for imaging biological samples. Such strategies provide a rich palette of light-molecule interactions that produce an optical signal from which an optical microscope image may be formed. In this special issue of the Journal of Biophotonics, we are celebrating the wide-ranging contributions that our dear colleague Gabi Popescu made to this field. Gabi was a big champion and cheerleader for this field and his enthusiasm for the widespread utility of label-free imaging was infectious.

There are a wide range of label-free imaging modalities. Each modality probes particular features of the specimen and each exhibits a sensitivity that depends on the sample properties and the experimental scenario. However, the field lacks a comprehensive comparison between various techniques to determine when each method will provide useful information, as well as an assessment of the detection sensitivity of these methods. In this work, we develop a general model for label-free signal generation to facilitate investigation of the relative performance of these label-free imaging methods.

Our analysis considers a universal light-matter interaction mechanism for label-free imaging techniques, then we apply the tools of statistical information theory to study the detection limits with label-free imaging methods. This strategy establishes bounds on the detection sensitivity of label-free microscopy. Note that we do not treat label-free methods based on the autofluorescent properties of a small set of endogenous biomolecules, as these methods cannot be incorporated into our general optical signal model. 

A wide range of label-free optical interactions have been exploited for optical microscopy. These optical modalities universally rely on optical spectroscopy of illumination light and the methods in which the light-matter interactions in the specimen modify light propagation, polarization, or color. Label-free imaging most often relies on linear optical scattering, where spatial variations in the optical susceptibility, $\delta \varepsilon$, distort light propagation through a specimen. To recover the three-dimensional variation in optical susceptibility, a range of optical methods can record quantitative changes in optical phase and amplitude and solve an inverse scattering problem. While such quantitative phase microscopy methods \cite{mir2012quantitative, jin2017tomographic, park2018quantitative} can be ubiquitously applied to specimens, the optical spectroscopy shows little dispersion, and as a result it is difficult to differentiate between particular molecular species \cite{park2009spectroscopic}. Nonlinear optical scattering processes of second- and third-harmonic generation (SHG, THG) can occur for a large incident optical field strength. These nonlinear scattering mechanisms convert incident light into a new color and reveal tissues formed from organized distributions of structural proteins (SHG) \cite{campagnola2011second, masihzadeh2010label, smith2013submillisecond, hu2020harmonic, james2021recent, Murray:23} or from morphologies such as cell membranes and small lipid bodies (THG) \cite{barad_nonlinear_1997, muller_3d_1998, debarre2006imaging, masihzadeh2009enhanced, masihzadeh2015third, weigelin2016third}.

The rise of label-free microscopy has facilitated our ability to chemically specify biochemical targets, allowing us to visualize cellular organelles without perturbing the biological dynamics. Because the identification and observation of the behavior of biomolecules provides critical insight into biological systems, methods that can provide label-free biochemical detection are highly sought after and forms the basis of a several label-free imaging methods that differentiate molecules based on their vibrational spectral fingerprints \cite{ chan2008raman, cheng2015vibrational, zhang2015coherent, wrobel2018infrared} or based on the excited state decay dynamics  \cite{ye2009nonlinear, stringari2015vivo, fischer2016invited, datta2020fluorescence}. The simplest vibrational spectral measurements exploit direct mid-infrared absorption at vibrational frequencies for which motion induces a change in the molecular dipole, thus producing direct optical absorption with incident light that matches the vibrational energy. \cite{stuart2006infrared, matthaus2008infrared, sabbatini2017infrared, shi2020mid} Alternatively, the Raman-active vibrational spectroscopy can be probed, where vibrational motion leads to a change in molecular polarizability and thus drives inelastic optical scattering, where scattered light either gains or loses a quanta of vibrational energy. \cite{ petry2003raman, dodo2022raman} Conventional Raman spectroscopy and imaging are limited in detection sensitivity because they rely on spontaneous Raman scattering, a rare process. Stimulated Raman scattering techniques, such as coherent anti-Stokes Raman scattering (CARS), \cite{alfonso2014biological, zhang2018perspective}, coherent Stokes Raman scattering (CSRS), \cite{ heuke2023coherent}, stimulated Raman scattering (SRS), \cite{lee2017imaging, rigneault2018tutorial}, or impulsive stimulated Raman scattering (ISRS) \cite{bartels2021low, wilson2008synthetic, wilson2008phase, schlup2009sensitive, domingue2014time, smith2022nearly, shivkumar2023selective, smith2024low} greatly increase the Raman signal scattering.

An advantage of stimulated spectroscopic interactions for the imaging of molecular targets is that the rate of signal generation may be elevated relative to the natural excited state relaxation times that constrain fluorescent imaging rates. The rate of signal generation can be increased in pump-probe experimental arrangements such as transient absorption (TA), excited state absorption (ESA), stimulated emission (SE), or  ground state depletion (GSD) \cite{ye2009nonlinear, min2011coherent, fischer2016invited} imaging methods. In this family of interactions, a pump pulse drives electronic absorption that perturbs the transmitted power of a time-delayed probe pulse. \cite{wong2021way}

Following excitation of a molecular chromophore, electrons promoted to an excited state will relax back down to the ground state and this excess energy is thermalized. Thermalization of the deposited energy heats the region surrounding the chromophore---producing an increase in temperature and pressure. These two perturbations are exploited for photoacoustic (PA) \cite{ li2009photoacoustic} and photothermal (PT) \cite{zharov2005photothermal, adhikari2020photothermal} detection mechanisms. We study the latter here because the detection is optical and thus relevant to our signal model. PT interactions can be driven by optical absorption \cite{cognet2008photothermal} or vibrational state transitions \cite{ zhang2019bond, bai2021bond, zhu2023stimulated} that leave residual energy in the molecule. The temperature change induced by the energy dissipation following optical excitation produces a small change in the effective linear optical susceptibility, $\delta \varepsilon$. This differential change  $\delta \varepsilon$ can then be extracted by comparing optical phase images or changes in optical scattering following excitation to those taken at thermal equilibrium. 

To link all these disparate label-free optical interactions together, we consider a description that can incorporate the signal model for each of these modalities. The signal model that we develop links all label-free imaging methods together to highlight the key underlying signal generation mechanism. To link all these disparate label-free optical interactions together and assess their relative detection limits, we employ the general signal model and compute the information available in measurements using statistical estimation theory. This model does allow direct comparison of the detection sensitivity between all the methods. Specifically, we consider scattering-induced changes in an optical imaging field produced by a spherical perturbation of optical susceptibility $\delta \varepsilon$. A model of the imaging field that has passed through an optical microscope is developed so that a model of the signal detection probability may be constructed. This signal model accounts for shot noise in the optical detection, capturing the limiting case of optical detection in the standard quantum limit. On the basis of this model, the measurement information is quantified by Fisher information and the fundamental limits in estimation precision for $\delta \varepsilon$ are assessed by the Cram\'er-Rao Lower Bound (CRLB). The effective susceptibility perturbation is then calculated for the label-free imaging methods presented in the introduction so that detection bounds of molecular concentration, or other parameters of interest, may be established with the general model. This general analysis can be applied to any label-free spectroscopy or imaging method and we hope will be a valuable tool for assessing label-free imaging experiments.


\section{\label{sec:theory} The imaging model for signal detection}

\begin{figure}
    \centering
    \includegraphics[width=0.9\textwidth]{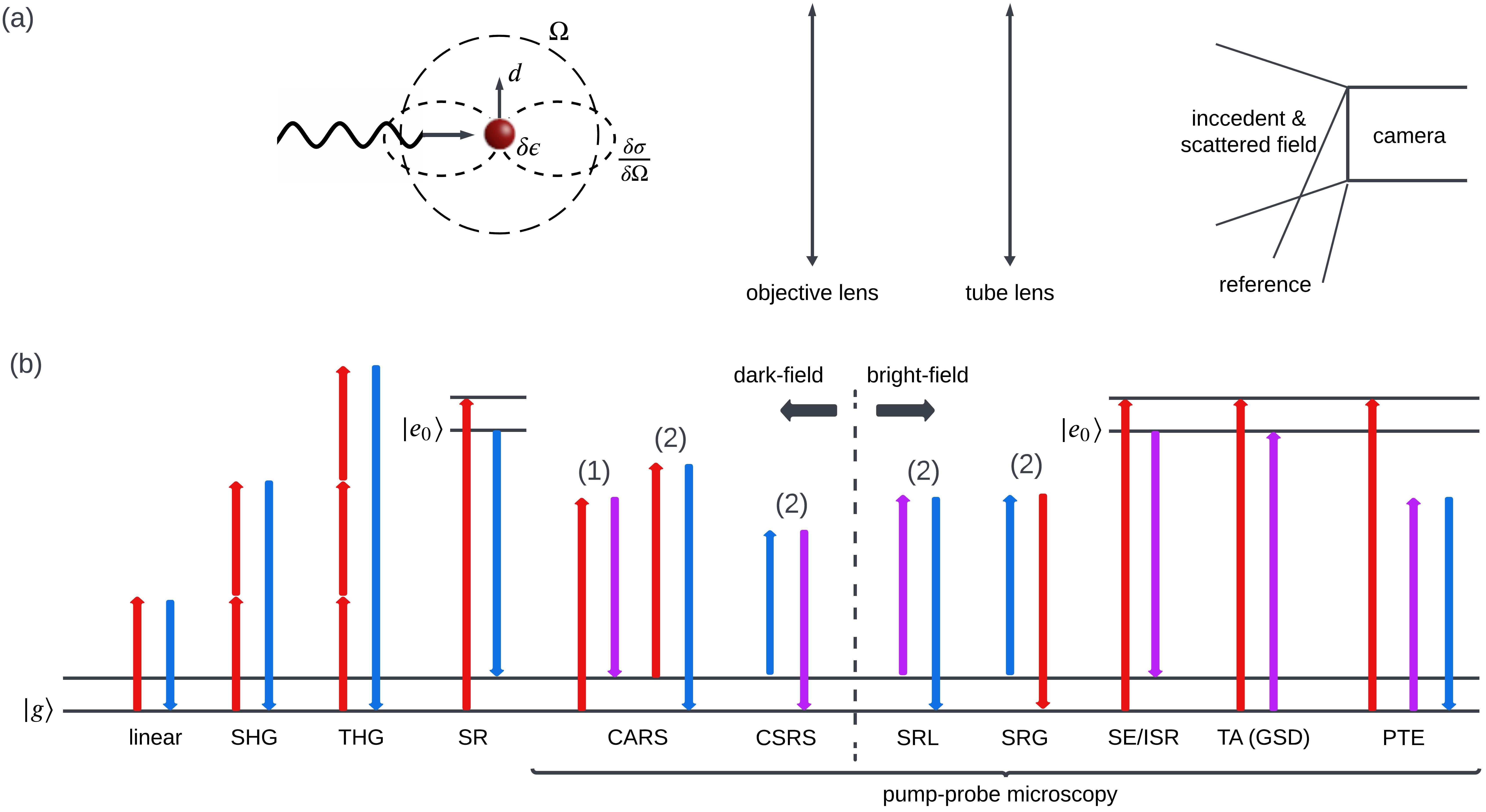}
    \caption{The model and the label-free imaging techniques. (a) From left to right: the scattering of the particle, the 4-f imaging configuration, and the measurement by a camera. In the scattering event, the particle is illuminated by an incident light, forming a dipole moment $\v d$. The scattering property of the dipole in the far-field region. (b) The contrast mechanisms of the imaging methods. The methods are classified into dark-field and bright-field methods. In the pump-probe methods, the red arrow denotes the pump, the purple arrow denotes the probe, and the blue arrow denotes the scattering, that is, the spatial frequency of the particle to be imaged~\cite{wang_three_2022}. Notice that CARS, CSRS, SRL, and SRG share the same step (1) so only step (2) is shown for CSRS, SRL, and SRG.}
    \label{fig:model}
\end{figure}

\noindent Our universal model for label-free signal generation is based on the optical system illustrated in Fig. \ref{fig:model}. We consider an object that consists of a spherical perturbation of optical susceptibility, $\delta \varepsilon = \varepsilon_s - \varepsilon_b$, a change in the relative dielectric permittivity of the sphere, $\varepsilon_s$, relative to a background relative dielectric permittivity, $\varepsilon_b$. The sphere has a radius $a \ll \lambda$ that is small compared to the incident field wavelength $\lambda$. The signal model is determined by the light that is imaged from the object space to the image space through a 4-f optical microscope; we closely follow the theoretical analysis of the image of a dipole in an optical microscope. \cite{khadir2019quantitative} Once the model for the signal is obtained, we apply the tools of statistical estimation theory to establish the bounds on the precision with which we may estimate $\delta \varepsilon$.

\subsection{The dipole moment of a sub-wavelength particle}

\noindent Our particle lies in an object space, with coordinates $\v r = (x,y,z)$ centered at the origin $\v r_0 = (0,0,0)$, and is illuminated by an incident optical field $\v E_{\rm i}(\v r, t) = E_0 \, u_r(\v r) \, u_t(t) \, \exp(i \, n_b \, k_0 \, z_0) \, \mathbf{\hat{\epsilon}}$. We assume that the incident beam propagates along $z$ so that the beam polarization vector, $ \mathbf{\hat{\epsilon}}$, lies in the transverse plane with a surface normal aligned with $z$. The incident beam has a peak electric field strength of $E_0$, with a spatial variation of field amplitude that varies due to beam diffraction described by $u_r(\v r)$. This diffracted field is normalized such that $|u_r| \le 1$. For cases of a pulse, we assume that $u_t$ is a complex temporal envelope normalized to $|u_t| \le 1$ and that the pulse is a member of a pulse train with a repetition rate of $\nu_r$. Note that for an un-pulsed continuous wave beam, $u_t = 1$. We assume that the illumination beam is centered at the origin such that $ u_r(\v r_0) =1$. The polarization of the incident beam is denoted by $\hat{\v \epsilon}$ and the free space wavenumber is $k_0 = 2 \, \pi / \lambda$ is increased by the object space background refractive index, $n_b = \sqrt{\varepsilon_b}$. The beam transverse intensity profile is $I(\v r_\perp, t) = I_0 \, |u_r(\v r_\perp, z)|^2 \, |u_t(t)|^2$ at an axial plane $z$, where the transverse spatial coordinate vector is $\v r_\perp  = (x,y)$. Definitions of the beam area, $A_b = I_0^{-1} \int  I(\v r_\perp, 0) \, d^2 \v r_\perp$, and the pulse temporal duration, $\tau_p = I_0^{-1} \int  I(0, t) \, d t$, make use of this transverse intensity profile. Finally, we note that the average power of the incident beam is $p_i = \nu_r  \, \int I_i(\v r_\perp, t)\, d t \, d^2 \v r_\perp = \nu_r  \, \tau_p \, A_b \, I_0 \equiv A_b \, I_a$, which can be written in a very simple form using the average intensity $I_a = \nu_r  \, \tau_p \, I_0$.

Due to constraints of optical diffraction, the beam area is on the order of $\lambda^2$ or larger. Consequently, scattering by the sub-wavelength centered on the illumination beam is produced by an oscillating electric dipole moment driven by the field at the peak of the incident beam, $\v \mu = \tilde{\alpha} \,  \v E_{\rm i}(\v r_0)$, where we assume that the particle is centered on the beam. The parameter $\tilde{\alpha} \equiv \varepsilon_0 \, \alpha$ is a complex quantity describing the propensity of a scatterer to produce a polarization in response to an applied electric field. We separate this polarizability into a product of the dielectric permittivity of free space, $\varepsilon_0$, and the complex-valued polarizability volume $\alpha$. Even for a real-valued optical susceptibility perturbation, $\delta \varepsilon$, the polarizability will be complex due to rescattering described by the radiative reaction term. \cite{carminati2021principles} However, because we consider $k_0 \, a \ll 1$ and $\delta \varepsilon \ll 1$, we may reliably approximate the polarizability volume by $\alpha \approx V \, \delta \varepsilon$, where the volume of the sphere is $V = (4 \, \pi/3) \, a^3$.

Within the approximations presented here, the oscillating dipole moment, $\v \mu$, radiates a dipole electric field. This radiation is the universal physical origin of our label-free signal used for probing a specimen. The frequency at which the induced dipole oscillates is determined by the particular label-free interaction and the interaction can either be an elastic scattering process, where the wavelength stays the same as the incident light, or an inelastic scattering process, where there is a change in optical wavelength. In addition, the effective susceptibility perturbation will be determined by the number of molecules contributing within the perturbation volume, $V$, and the coherence of the interaction that drives the susceptibility perturbation. The specific cases for label-free imaging will be discussed later in this article.

\subsection{Dipole scattering and cross-sections}

\noindent To appreciate signal levels, we consider the expression for the extinction, scattering, and absorption cross-sections of the label-free perturbation. \cite{myroshnychenko2008modelling} The extinction cross-section reads $\sigma_{\rm ext} = (k_0/n_b) \, V \, \mathrm{Im} \{ \delta \varepsilon \}$, while the scattering cross-section is given by $\sigma_{s} = (k_0^4 \, V^2/ 6 \, \pi) \, |\delta \varepsilon|^2$. The difference is the absorption cross-section, $\sigma_{\rm abs} = \sigma_{\rm ext} - \sigma_s$. Note that these cross-section values will be composed of molecular components within our scattering sphere and the contribution of the effective polarizability of the set of molecular components to the total $\delta \varepsilon$.  As a consequence, the total cross-section values depend on the coherence between induced dipole moment between each molecules within the sphere. Each scenario will be treated separately later in the article. 

With the cross-section for the sub-wavelength sphere, we can estimate the extinguished and scattered average powers as $p_\sigma = \nu_r \, \sigma \int I_i(0, t) \, d t$. Making use of the expression for the incident average power, we write $p_\sigma = (\sigma/A_b) \, p_i$, scaling as the ratio of the cross-section and the incident beam area. For a fixed incident optical power budget, it is desirable to focus the incident beam tightly to minimize $A_b$, and thus maximize the intensity of the illumination beam. In a high numerical aperture (NA) focusing limit, an incident beam focused to the origin can be approximated with a three-dimensional Gaussian distribution $u_r(\v r) = \exp(-\rho^2/w_0^2) \, \exp(- z^2/w_z^2)$. Here $\rho = \sqrt{x^2 + y^2}$ is radial transverse coordinate and the radii of the focused beam are $w_0 = 0.52 \, n_b \, \lambda /\mathrm{NA}$ and $w_z = 0.76 \, \lambda /[1-\sqrt{1-(\mathrm{NA}/n_b)^2}]$. \cite{mertz2019introduction} This produces a beam area of $A_b = 0.42 (n_b/\mathrm{NA}) \, \lambda^2$. Combining these expressions, we find that the power (scattered, extinguished, or absorbed) reads $p_\sigma \approx 2.4 \, (\mathrm{NA}/n_b)^2 \, (\sigma/\lambda^2) \, p_i$. The coefficient in front of the incident power is a small number indicating that much of the incident power is unperturbed. This sets a level of background light that usually degrades sensitivity and must be accounted for in the signal model. For a complete model, we compute the image of excitation beam and scattered beam through a 4-f imaging system.


\subsection{Signal obtained from the image of the dipole through a microscope}

To compute the signal produced from a field imaged from the object region into an image region, we follow the derivation of the image of a dipole field produced with a high NA imaging system. \cite{khadir2019quantitative} The object field may be expanded into a set of transverse spatial frequencies, $\v k_\perp = (k_x, k_y)$, at a reference plane, here $z=0$, providing the expression of the object field
\begin{equation}
\label{eq:E_obj}
\v E_{\rm obj}(\v r) = \int \v e(\v k_\perp) \, e^{i \, (\v k_\perp \cdot \v r_\perp + \gamma_o \, z)}d^2 \v k_\perp.
\end{equation}
The transverse spatial frequency spectral amplitude is denoted by $\v e(\v k_\perp)$ and the angular spectral propagator can be used to express the field at a plane other then $z=0$ with a transverse spatial frequency phase determined by the axial spatial frequency for propagation, $\gamma_0 = \sqrt{n_b^2 \, k_0^2 - \lVert \v k_\perp \rVert^2}$ of the wavevector, at $\v k = (\v k_\perp,\gamma_o)$, for each transverse spatial frequency. 

The object field is mapped to an image space where we place a detector using a 4-f optical imaging system. As indicated in Fig. \ref{fig:model}, the focal lengths of the imaging system are $f_1$ and $f_2$, leading to a magnification of the object field by the factor $M = -f_2/f_1$. When mapping the object field to the image space in air, the wavevector is transformed such that $\v k' = (\v k_\perp', \gamma_i)$, where $\v k_\perp' =(k_x/M,k_y/M)$ and $\gamma_i = \sqrt{k_0^2 - \lVert \v k_\perp' \rVert^2 }$. The image of the object field, labeled by the image-space coordinates $\v r_1 = (\v r_{1 \perp}, z_1)$, at the in-focus plane of $z_1 = 0$ may be expressed in terms of the object space transverse spatial frequencies as
\begin{equation}
    \label{eq:E_img}
\v E_{\rm img}(\v r_1) = M \, \int \sqrt{\frac{\gamma_o}{\gamma_i}} \, \mathcal{H}(M \, \v k_\perp') \, \v e_1(M \, \v k_\perp') \, e^{i \, \v k_\perp' \cdot \v r_{1 \perp} }d^2 \v k_\perp'.
\end{equation}
Moreover, the polarization of the dipole field may be decomposed into $\hat{s}$ and $\hat{p}$ polarization directions. The imaging system transforms the object field polarization through a unitary rotation, thus preserving the magnitude of the transverse spatial frequency components, so that $|\v e_1| = |\v e|$. The coherent transfer function of the 4-f imaging system, $\mathcal{H}(\v k_\perp)$, is a low-pass transverse spatial frequency filter with a cutoff spatial frequency, $k_c = 2 \, \pi \, \mathrm{NA}/\lambda$, determined by the NA of the imaging objective lens ($f_1$).

We assume that the NA of the imaging objective lens does not restrict the collection of the illumination beam, which implies the beam is simply expanded by the magnification factor with a corresponding drop in field amplitude such that the full power of the incident excitation beam is transmitted through the imaging system. The label-free signal originates from the source, $\v Q = (k_0^2/\varepsilon_0) \, \mathbf{\mu}$, that is produced by the induced dipole moment. The transverse spatial frequency spectrum of the forward-propagating component of the dipole field reads
\begin{equation}
\label{eq:eQ}
\v e_{\rm Q}(\v k_\perp) = \frac{i}{8 \, \pi^2 \, \gamma_o}[\v Q - (\v Q \cdot \hat{\v k})\hat{\v k}].
\end{equation}
Here $\hat{\v k} = \v k / \lVert \v k \rVert$ and we exploited that fact that we consider an object is located at the origin of the coordinate system. The power of the field scattered in the forward direction is given by \cite{khadir2019quantitative} the formula $p_{\rm sob} = (2 \, \pi^2/ \mu_0 \, \omega) \int |\v e(\v k_\perp)|^2 \, \gamma_o \, d^2 \v k_\perp$. Applying this to the dipole scattered field, we find that $p_{\rm sob} = \sigma_s \, I_0/2$, which is half of the total scattered power. The other half of the power propagates in the backward direction. Due to the finite NA of the imaging objective lens, the power of the image of the dipole emission is reduced to, $p_{\rm sim} = \eta_c \, p_{\rm sob}$,where the efficiency of the forward-scattered power by the object is given by $\eta_c = 1- \sqrt{1 - \mathrm{NA}^2} + (\mathrm{NA}^2/4) \, \sqrt{1 - \mathrm{NA}^2}$.

To admit a wide range of experimental arrangements, we consider the following total field in the image space
\begin{equation}
\label{eq:E_tot}
    \v E_{\rm tot,im} = a \, \v E_{\rm ex,im} + \v E_{\rm Q, im} + \v E_{\rm r}  \equiv \tilde{r} \, \v E_{\rm ex,im} + \v E_{\rm Q, im}.
\end{equation}
This total field consists of the dipole field image, $\v E_{\rm Q, im}$, the image of the excitation field, $\v E_{\rm ex,im}$, and a reference field $\v E_{\rm r} = r \, \v E_{\rm ex,im}$. We have assumed that the reference field is a replica of the image of the illumination field, scaled by a complex factor $r = R \, \exp(i \, \phi)$, where $R$ is the relative amplitude and $\phi$ is the relative phase of the reference beam. In addition, we multiply a factor $a$ by the image of the excitation field so that we can set $a = 1$ to represent bright-field imaging and $a = 0$ for the case of dark-field imaging, where the unscattered field is not collected. We define a generalized complex reference amplitude as $\tilde{r} = a + r$.

In our Fisher information analysis, we consider both the case where a single detector collects some fraction of the total signal power or we send the image onto a camera to capture the signal. In both cases, the information content is the same as the measurement on each pixel is uncorrelated. As a result, the relevant parameter from the signal model starts with the total power of the signal, which is computed from the total transverse spatial frequency amplitude. This total power is the sum of three terms, $p_t = |\tilde{r}|^2 \, p_i + p_{\rm sig} + p_{\rm int}$. The first two power terms have already been computed and the interference power term, $p_{\rm int} = (4 \, \pi^2 \, \tilde{r}^*/ \mu_0 \, \omega) \int |\mathcal{H}(\v k_\perp)|^2 \, \mathrm{Re} \{  \v e_{Q}(\v k_\perp) \cdot \v e_{\rm ex}^*(\v k_\perp) \} \, \gamma_o \, d^2 \v k_\perp$, arises from mixing between the dipole source transverse spatial frequency distribution given in \reqn{eQ} and the excitation beam transverse spatial frequency distribution, $\v e_{\rm ex}(\v k_\perp) = (4 \, \pi^2)^{-1} \, \int u(\v r_\perp) \, \exp(- i \, \v k_\perp \cdot \v r_\perp) \, d^2 \v r_\perp$. Assuming that we have a symmetric, unaberrated beam propagating along the optical axis that was produced by a uniformly filled illumination optic numerical aperture ($\mathrm{NA}_i \le \mathrm{NA}$), then we find that $p_{\rm int} = - I_0 \, k_0 \, V \,  \mathrm{Im} \{ \tilde{r}^* \, \delta \varepsilon \} \, f_{\rm NA}$, where $f_{\rm NA} = 1 - (\mathrm{NA}_i/2 \, n_b)^2$. Note that plane wave illumination is the limiting case where $\mathrm{NA}_i \rightarrow 0$.

The signal collected in a time-interval $\Delta t$, for a photon energy $\mathcal{E}_{\rm ph} = h \, c /\lambda$ and detector quantum efficiency $\eta_d$ that has a surface area larger than the beam size gives us a mean-detected photon count given by $\mathcal{N} = \eta_d \, \Delta t \, p_t /\mathcal{E}_{\rm ph}$, reads
\begin{equation}
\label{eq:n}
\mathcal{N} = \mathcal{N}_i  \, \left( |\tilde{r}|^2 + \frac{1}{2} \, \eta_c \, \frac{\sigma_{\rm eff}^{(j)}}{A_b} - \frac{k_0 \, V}{A_b} \, \mathrm{Im} \{ \tilde{r}^* \delta \varepsilon_{\rm eff}^{(j)} \} \, \ell^{(j)}  \, f_{\rm NA} \right) \equiv \mathcal{N}_i  \, F_N.
\end{equation}
Where we have defined $\mathcal{N}_i = \eta_d \, p_i /\mathcal{E}_{\rm ph}$ the mean number of detectable photons in the illumination beam and the single-photon signal function $F_N$. In addition, we have defined $\delta\varepsilon_{\rm eff}^{(j)} = N \, B^{(j)} \, \alpha^{(j)}$ for use with coherent nonlinear scattering as the effective susceptibility becomes dependent on the incident fundamental beam power, and of course, with CARS and CSRS, will depend on Stokes and pump beam powers. The parameter $B^{(j)}$ and $\ell^{(j)}$ will be defined when coherent nonlinear scattering is discussed and account for pulse averaging effects. For the linear case, where $j=1$, $B^{(1)} = 1$ and $\ell^{(1)} =1$. In addition, $\sigma_{\rm eff} = k_j^4 \, V^2 \, |\delta\varepsilon_{\rm eff} |^2/6 \, \pi$ defines the effective scattering cross-section. The number density, $N$, of the molecules in the sphere leads to a total number of molecules of $N \, V$, each with a single-molecule polarizablity (for $j=1$, where $\alpha^{(1)}$ is the molecule polarzaiblity) and for $j>1$ we represent hyperpolarizabilities used to describe nonlinear scattering processes.

\subsection{Probabilistic Model for Susceptibility Perturbation Estimation and CRLB}\label{sec:prob_crlb}

Having established the measurement model in terms of photon count for the particle susceptibility in label-free imaging, we now delve into the quantitative assessment of each imaging system's sensitivity of the measurement data, or equivalently, the amount of information the measurement data carries about the particle susceptibility. The statistical tools used are the Fisher information, $J$, and the CRLB, both of which are instrumental in quantifying the fundamental limits in estimation precision. 

When only photon detection noise is present, noise in optical detection can be modeled with as a Poisson process, where the likelihood function for detecting $Y=y$ photons to be detected is expressed as $f(Y=y;\delta\varepsilon_{\rm eff})=\mathcal{N}^{y}\,e^{-\mathcal{N}}/y!$ and $\mathcal{N}$ is the mean photon count. Here we analyze the estimation precision of $\delta \varepsilon$ in the model given in \reqn{n} for a Poisson noise model. It is beneficial to define a normalized Fisher information $\tilde{J}$ by the number of photons from the incident light $\mathcal{N}_{i}$ since the Fisher information scales linearly with respect to the signal strength. The stronger the detected signal, which requires an increased illumination power, the larger the Fisher information. The normalized Fisher information $\tilde{J}$ signifies the amount of information carried by a single incident photon about the object susceptibility perturbation. The Fisher information for this estimate, $J = \mathcal{N}_i \, \tilde{J}$, can be separated into a product of the incident mean photon count $\mathcal{N}_i$ and the normalized single-photon Fisher information, $\tilde{J}$, that provides information on the sensitivity for an experimental arrangement on the detection of the parameter of interest. The Fisher information and the CRLB are inherently connected as the CRLB is inversely proportional to the Fisher information. Serving as a theoretical lower limit on the variance of any unbiased estimator when evaluated at the true parameter value, the CRLB for any unbiased estimation of the susceptibility is, therefore, given by $\sigma_{\mathrm{CRLB}}^{2}=J^{-1}=\mathcal{N}_{i}^{-1}\tilde{J}^{-1}$. The limit to the precision with which a single parameter is simply $\sigma_{\rm CRLB} = 1/\sqrt{J}$. Multiparameter estimation is more complex because the Fisher information becomes a matrix which must be inverted to obtain the CRLB values for estimation precision on its diagonal.

For the Poisson noise model, the normalized Fisher information, $\tilde{J} = s_N^2 / F_N$, is the ratio of the square of the single photon Fisher score, $s_N$, to the single-photon signal flux $F_N$. The Fisher score, which is the derivative of the log likelihood function for the measurement with respect to the parameter of interest, establishes the sensitivity of the measurement with respect to the parameter of interest and helps quantify the amount of information that a set of data provides about the parameter of interest in a statistical model. Because in general $\delta \varepsilon$ is a complex-valued parameter, we will consider the limiting cases where our parameter of interest, $\delta \varepsilon$, is either purely real so that $\delta \varepsilon \rightarrow \delta \varepsilon_{\rm re}$ or it is purely imaginary so that $\delta \varepsilon \rightarrow i \, \delta \varepsilon_{\rm im}$. Then, we consider two cases: the normalized Fisher information for a real-valued susceptibility perturbation is given by $\tilde{J}_{\rm re} = (s^{\rm (re)}_N)^2/F_N^{\rm (re)}$ and the normalized Fisher information for an imaginary susceptibility perturbation is given by $\tilde{J}_{\rm im} = (s^{\rm (im)}_N)^2/F_N^{\rm (im)}$. For the real-valued case, we use
\begin{equation}
    F_N^{\rm (re)} = |\tilde{r}|^2 + \, \eta_c \,  \frac{k_0^4 \, V^2}{12 \, \pi \, A_b} \, \left(\delta \varepsilon_{\rm eff}^{\rm (re)}\right)^2 - \frac{k_0 \, V}{A_b} \,  \delta \varepsilon_{\rm eff}^{\rm (re)}\, \mathrm{Im} \{ \tilde{r}^* \} \, \ell \, f_{\rm NA}
\end{equation}
and
\begin{equation}
    s_N^{\rm (re)} = \eta_c \, \frac{k_0^4 \, V^2}{6 \, \pi \, A_b} \, \delta \varepsilon_{\rm eff}^{\rm (re)}  - \frac{k_0 \, V}{A_b} \, \mathrm{Im} \{ \tilde{r}^* \} \, \ell \, f_{\rm NA}.
\end{equation}
Similarly, for the imaginary-valued case, we use
\begin{equation}
    F_N^{\rm (im)} = |\tilde{r}|^2 + \, \eta_c \,  \frac{k_0^4 \, V^2}{12 \, \pi \, A_b} \, \left(\delta \varepsilon_{\rm eff}^{\rm (im)}\right)^2 - \frac{k_0 \, V}{A_b} \,  \delta \varepsilon_{\rm eff}^{\rm (im)}\, \mathrm{Re} \{ \tilde{r}^* \} \, \ell \, f_{\rm NA}
\end{equation}
and
\begin{equation}
    s_N^{\rm (im)} = \eta_c \, \frac{k_0^4 \, V^2}{6 \, \pi \, A_b} \, \delta \varepsilon_{\rm eff}^{\rm (im)}  - \frac{k_0 \, V}{A_b} \, \mathrm{Re} \{ \tilde{r}^* \} \,  \ell \,f_{\rm NA}.
\end{equation}
The normalized Fisher information will be explored for a variety of label-free signal detection modalities and experimental arrangements.

The Fisher information and the CRLB can be connected to the notion of signal and noise, and thus the signal-to-noise ratio (SNR) that are commonly used to describe optics experiments. The change in the expected mean signal for the true parameter of interest value $\delta \epsilon_{{\rm eff},0}$ to be measured is approximated by $\Delta \mathcal{N}_s \approx \mathcal{N}_i \, s_N \, \Delta \delta \epsilon_{\rm eff}$ given that $\Delta \delta \epsilon_{\rm eff}$ is sufficiently small. Equivalently, when $\delta \epsilon_0$ is sufficiently small, the expected mean signal is given by $\mathcal{N}_{s}\approx \mathcal{N}_{i}\,s_{N}\,\delta \epsilon_{{\rm eff},0}$. The root mean square noise for the Poisson noise model here gives a noise of $\mathcal{N}_n = \sqrt{\mathcal{N}_i\,F_{N}}$. From these quantities, we may construct the SNR as $\mathrm{SNR} \equiv \mathcal{N}_s/ \mathcal{N}_n$. This leads to the understanding of how the Fisher information and the CRLB relates to the balance between the signal and the noise. If a small change in the parameter of interest results in a larger change in the signal or the noise is lower, the Fisher information increases and the CRLB decreases.  Comparing this definition, we see that we may write  $\mathrm{SNR} = \delta \varepsilon_0 \, J^{1/2} = \delta \varepsilon_0 / \sigma_{\rm CRLB}$. This suggests that the SNR can be comprehensively represented by the behavior of the maximum likelihood estimator (MLE) as it follows a Gaussian distribution with its mean equal to the true parameter value, and its variance approaches the CRLB when the number of measurements approaches infinity. Essentially, the variability in the MLE relative to the mean from different realizations of the same process is contingent on the level of noise power. A lower noise power results in diminished relative variability, highlighting the importance of noise control. Moreover, We may interpret the limit of detection as the object susceptibility perturbation at which the SNR $=1$ in our measurement and therefore, $\left(\delta\varepsilon_{{\rm eff},0}\right)_{\rm min} = \sigma_{\rm CRLB}$.

In summary, a higher Fisher information implies a lower CRLB, indicating that precise estimation of the parameter is attainable. Therefore, our following sensitivity analysis will focus on the calculation of the Fisher information for all label-free microscopy methods. Moreover, the Fisher information is broader than a calculation of the SNR and significant difference can emerge between a simple SNR analysis and one based on Fisher information. \cite{barrett2007maximum, knee2015fisher} While SNR analysis is applied to a measure of total signal and total noise, \cite{smith2013submillisecond} it is not straightforward to apply such analysis to a multipixel detector such as a camera. \cite{dong2021fundamental} The Fisher information analysis presented here is directly applicable to detection schemes where a fraction, $\eta_c/2$, of signal power is collected on a single photodetector, or where a camera is used and many pixels are used to collect the data over an image field of view. At first glance, these may not seem compatible; however, each pixel measurement is an independent event and thus the log likelihood functions for each pixel add together, meaning that Fisher information will add for each pixel and becomes equivalent to the integrals used in our analysis and thus our results are equally valid for camera-based detection. \cite{khadir2020full} 


\section{Effective Susceptibility of Label-free Optical Interactions}\label{sec:eff_sus}

The normalized Fisher information, $\tilde{J}$, provides the contribution to the standard deviation of susceptibility perturbation estimation precision, $\tilde{J}^{-1/2}$, which we have seen also corresponds to the minimum detectable perturbation for a general scattering model. Each label-free interaction produces a susceptibility perturbation that is related to the intensity of the light that drives the light-matter interaction and sets bounds on the limit on how small of a concentration of the molecules under study may be detected. The coherence of the scattered light relative to the incident light also plays a role the effective $\delta \varepsilon_{\rm eff}$ and the detection limits. In all cases, we consider a set of molecules contained within the scattering sphere volume, $V$, that produce a total detected light power. 

We may separate the label-free optical interactions into two broad categories: coherent and incoherent. In the incoherent case, each molecule contributes to the change in the detected light power independently of the other molecules within the interaction volume. The total signal is proportional to the concentration in the incoherent case. In the coherent case, each molecule is driven in phase within the interaction volume, and thus the total susceptibility perturbation is proportional to the molecular concentration, which we will specify in terms of the number density, $N$. Within the volume, there are $M = N \, V$ total molecules that contribute to the label-free signal generation.



A key aspect of the incoherent case is that the phase of scattering or emission from each molecule fluctuates randomly on a time scale that is rapid compared to the detector integration time. This is the case for autofluorescence and for spontaneous Raman scattering. In spontaneous Raman scattering, each molecule scatters light inelastically to new optical frequencies by modulation of the molecular polarizability due to thermal excitation of molecular vibrational modes. The phase of the vibrational oscillations, and thus the phase of the scattered light, is a random variation that changes from molecule to molecule. Within $V$, each molecule will scatter a power of $p_R^{(1)} = \sigma_R^{(1)} \, I_0$. The origin of Raman scattering is a change in polarizablity, $\delta \alpha^{(1)} = \alpha' \, Q_v$, of the molecule with displacement of the vibrational coordinate, $Q_v$. As weak excitation of a vibrational mode is modeled as a harmonic with an amplitude of $Q_{v0} = \sqrt{\hbar/ 2\, \Omega_v}$ for vibrational frequency $\Omega_v$ driven by thermal excitation. The strength of the polarizability modulation is $\tilde{\alpha}'$. The Raman scattering cross-section of a single molecule follows as $\sigma_R^{(1)} = (k_0^4/6 \, \pi) \, |\delta \alpha^{(1)}|^2$. This classically derived model must be slightly modified to account for mode occupancy in a quantum scattering picture, and this modification explains the discrepancy between the amplitude of Stokes and anti-Stokes spontaneous Raman scattering.

Because spontaneous Raman scattering is incoherent, the total power scattered is simply $p_R = M \, p_R^{(1)}$. The effective Raman scattering cross-section for the volume is $\sigma_R^{V} = M \, \sigma_R^{(1)}$. In the case of Raman scattering without resonant enhancement, the effective Raman susceptibility perturbation is purely real, $\delta\varepsilon^{(\rm SR)} = k_0^{-2} \, \sqrt{6 \, \pi \, M \, \sigma_R^{(1)}} /V=\delta \varepsilon_{\rm eff}^{(\rm SR)}$. Raman scattering interactions are weak, which is reflected in low Raman scattering cross-sections ranging from $\sigma_R^{(1)} \sim 10^{-31} - 10^{-29} \, \mathrm{cm}^2$.\cite{gao2023absolute} Tuning the Raman laser near an electronic absorption resonance can increase the cross-sections to $\sigma_R^{(1)} \sim 10^{-25} - 10^{-21} \, \mathrm{cm}^2$. Thus, while Raman vibrational spectra are extremely valuable, detection at low species concentrations is exceedingly difficult and stimulated Raman and field enhancement techniques have been used to help alleviate this difficulty.


Another class of inelastic scattering processes are those of coherent nonlinear scattering where light at a fundamental frequency $\omega$ is incident on a molecule. If the amplitude of the incident field is sufficiently large, the induced dipole moment no longer exhibits a linearly proportional response to the applied electric field. This dipole moment is usually expanded as a Taylor series of the form \cite{brasselet2011polarization} $\v \mu = \tilde{\alpha} \, E + \tilde{\beta} \, E^2 + \tilde{\gamma} \, E^3  + \dots $. The quantities $\beta = \tilde{\beta}/\varepsilon_0$ and $\gamma = \tilde{\gamma}/\varepsilon_0$ are called the first and second hyperpolarizabilities. Here, we are assuming that the polarizability and hyperpolarizabilities are isotropic so that the complications of tensor algebra need not be invoked. We will use a compact notation by introducing $\alpha^{(j)}$ as a  generalized hyperpolarizability. Thus, we may write the induced nonlinear dipole as $\v \mu = \varepsilon_0 \sum_j \alpha^{(j)} \, E^j$. The second-order therm with $j=2$ can represent SHG,  where $\alpha^{(2)} = \beta$ is the hyperpolarizability. In the case of $j=3$, $\alpha^{(3)} =  \gamma$ is the second hyperpolarizablity, which includes the case of THG and self phase modulation driven by the electronic contribution, $\gamma_e$, and includes stimulated Raman scattering that arises from the use the vibrationally resonant component, $\alpha^{(3)} = \gamma_v$.

These hyperpolarizablities produce a nonlinear source term, $\v Q^{(j)} = k_j^2 \, \alpha^{(j)} \, E_i^j \,  \hat{\v \epsilon}$. Here the wavenumber at the scattered frequency is $k_j = \omega_j \, k_0$ and the harmonic frequency is $\omega_j = j \, \omega_0$. For a set of molecules in the volume, $V$, at a number density $N$, the coherent scattering is described by a nonlinear polarization density, $P^{\rm (NL)} = \varepsilon_0 \, D^{(j)} \, \chi^{(j)} \, E_i^j $. The factor $D^{(j)}$ is a degeneracy parameter with a value determined by the nonlinear interaction. In the volume of our sub-wavelength sphere with a number density $N$ of molecules, the generalized hyperpolarizabilities for SHG and THG are $\alpha^{(2)} \rightarrow \beta = \chi^{(2)}/2 \, N$ and $\alpha^{(3)} \rightarrow \gamma = \chi^{(3)}/4 \, N$, respectively. The nonlinear scattering cross-section, $\sigma_s^{(j)} = (2^{j-1} \, k_j^4/6 \, \pi \, n_b^j \, (\varepsilon_0 \, c)^{j-1}) \,  |N \, V \, \alpha^{(j)}|^2$, is defined through the instantaneous scattered power. As nonlinear optical interaction strengths are weak, pulsed lasers are used to ensure a large enough peak field strength to produce sufficient rates of nonlinear scattering.

The total time-averaged power scattered by a nonlinear dipole source with frequency $\omega_j$, whether from a single molecule or a distribution inside of a sub-wavelength sphere, is $p_j = \sigma_{\rm eff}^{(j)} \, I_a$. We have defined an effective linear cross-section for the nonlinear scattering process as $\sigma_{\rm eff}^{(j)} = \sigma_s^{(j)} \, g^{(j)} \, I_a^{j-1}$. Notice that this effective cross-section depends nonlinearly on the average intensity of the incident fundamental beam, $I_a$, and on the zero-lag $j^{\rm th}$-order intensity correlation function, $g^{(j)}$. This correlation function, defined as, $g^{(j)} =\langle I^j(t) \rangle /\langle I(t) \rangle^j$, depends on the duty cycle, $\nu_r \, \tau_p$, of the fundamental excitation beam laser source. While the exact value of $g^{(j)}$ depends on the pulse shape, the value is bounded by $g^{(j)} \le (\nu_r \, \tau_p)^{-(j-1)}$, where the upper bound is met with a square pulse. The effective cross-section defined an effective linear susceptibility perturbation through the relationship $\delta \varepsilon_{\rm eff} = \sqrt{6 \, \pi \, \sigma_{\rm eff}^{(j)} }/k_j^2 \, V$. With this, we define the effective linear susceptibility perturbation for coherent harmonic scattering as $\delta \varepsilon_{\rm eff} =N \, \alpha^{(j)} \, B^{(j)}$. The factor $B^{(j)} = \sqrt{(2/\varepsilon_0 \, c)^{(j-1)} \, g^{(j)} \, I_a^{(j-1)}}$ accounts for averaging the nonlinear scattered power over the pulse train. The effective cross-section and susceptibility can be used direction in the signal model given in \reqn{n}. This equation also includes the term $\ell^{(j)} = h^{(j)}/\sqrt{g^{(j)}}$. The term $h^{(j)} = I_a^{-(j+1)/2} \, \langle I(t)^{(j+1)/2} \rangle$ arises as an interference factor from signal averaging over the pulse train. This term is also bounded as $h^{(j)} \le \sqrt{g^{(j)}}$, where the bounds are again saturated by a square pulse. In the case of a square pulse, $\ell^{(j)}=1$.

Single molecules interacting can absorb light through linear or nonlinear absorption processes and optical absorption can occur for electronic and vibrational energy level transitions. Exactly on resonance, the polarizability for a molecule becomes purely imaginary, i.e., $\alpha^{(1)} \rightarrow i \, \alpha^{(1)}_i$. The superscript indicates that we are dealing with the polarizability of a single molecule. This polarizability produces both absorption and scattering, with cross-section for extinction, $\sigma^{(1)}_e = k_0 \, \mathrm{Im}\{ \alpha^{(1)} \}$, scattering, $\sigma^{(1)}_s = (k_0^4/6 \, \pi) |\alpha^{(1)}|^2$, and absorption, $\sigma^{(1)}_a  = \sigma^{(1)}_e - \sigma^{(1)}_s$. As generally $\sigma^{(1)}_a \ll \lambda^2$, the absorption cross-section on resonance, when $\alpha^{(1)}$ is purely imaginary, the absorption cross-section is well approximated by $\sigma^{(1)}_a \approx k_0 \,  \alpha^{(1)}_i$. The perturbation to the linear susceptibility form a number density, $N$, molecules is then $\delta \varepsilon_{\rm eff}^{\rm (abs)} = i \, N \, \alpha^{(1)}_i \approx i \, N \, \sigma^{(1)}_a / k_0$. The absorption cross-sections vary over a wide range, with a maximum value determined of $\sigma_a \sim \lambda^2/2$, which corresponds to $\delta \varepsilon_{\rm abs} = i \, N \, \alpha^{(1)}_i \approx i \, N \, \lambda^3/4 \, \pi$. Chromophores have absorption cross-sections ranging from \cite{gao2023absolute} $\sigma^{(1)}_a \sim 10^{-17} - 10^{-15} \, \mathrm{cm}^2$ for visible and ultraviolet absorption. These numbers drop several orders of magnitude for mid-infrared vibrational spectra that exhibit cross-sections in a range $\sigma^{(1)}_a \sim 10^{-19} -   10^{-17} \, \mathrm{cm}^2$.  Overtone stretches are generally weaker, on the order of range $\sigma^{(1)}_a \sim 10^{-22}$. \cite{cias2007absorption, wang2013spectroscopic} 

Another common absorption mechanism is multiphoton absorption, were promotion of an electron from a ground to an excited state requires the simultaneous arrival of two or more photons with energy below the energy gap. For degenerate two-photon absorption, the interaction of the fields induces a perturbation to the effective linear optical susceptibility of \cite{rumi2010two} $\delta \varepsilon_{\rm 2PA} = (3/4) \, \chi^{(3)} \, |E_i|^2$. This perturbation is complex-valued, indicating that both self-phase modulation and two photon absorption are driven in this interaction. Moreover,  the existing linear susceptibility dominates this interaction, i.e., $\chi^{(1)} \gg \delta \varepsilon_{\rm 2PA}$, and thus the change in field strength and the extinguished power is vanishingly small compared to two-photon absorption. As a result, two-photon and multiphoton absorption in general are typically used with efficient fluorophores, where emitted fluorescent light is collected as the signal. Thus, we do not discuss direct detection of molecules through multiphoton absorption in the context of direct detection. 

The limitations of spontaneous Raman scattering can be partially mitigated using stimulated Raman methods. These techniques are nonlinear optical methods where a two-photon resonant excitation is driven at the vibrational frequency, $\Omega_v$, in a molecule. The stimulated two-photon process driven by two incident fields, a pump field, $E_p$, at frequency $\omega_p$ and a Stokes field, $E_S$, at frequency $\omega_S<\omega_p$.  At resonance, the frequency difference is set to $\omega_p - \omega_S = \Omega_v$. There are many subtleties in dealing with the description of stimulated Raman scattering and we will focus on the vibrationally resonant part of the nonlinear optical response arising from $\chi^{(3)}_{\rm VR}$. However, the presence of nonlinear phase modulation from the electronic contribution to the nonlinear optical susceptibility presents challenges and opportunities---depending on the experimental arrangement. 

The CARS and CSRS nonlinear scattering processes also coherently produce light at a new optical frequency of $\omega_{\rm aS/cS}$, where aS and cS denotes the anti-Stokes and Stokes frequencies, respectively. As with SHG and THG, we define an effective susceptibility that may be computed in an analogous manner to the case of SHG and THG scattering, resulting in $\delta \epsilon_{\rm eff}^{\rm (CARS/CSRS)} = N \, B^{\rm (CARS/CSRS)} \, \gamma^{(CRS)}$. Here $B^{\rm (CARS/CSRS)} = (2/\varepsilon_0 \, c) \, \sqrt{g^{\rm (CARS/CSRS)} \, I_{\rm ap} \, I_{\rm aS}}$, which depends on the product of the average power of the pump and Stokes beams. Here we have also assumed that the temporal profile of the pump and Stokes pulses are identical so that $g^{\rm (CARS/CSRS)} =g^{(3)}$. The same is true of $h^{\rm (CARS/CSRS)} =h^{(3)}$. The explicit expression for the effective scattering cross-section now reads $\sigma_{\rm eff}^{\rm (CARS/CSRS)} = (2 \, k_{\rm aS/S}^4 \, V^2 /3 \, \pi \, n_b^3 \, (\varepsilon_0 \, c)^2) \,  |\delta \epsilon_{\rm eff}^{\rm (CARS/CSRS)}|^2$. The wavenumbers are given by $k_{\rm aS/cS} = \omega_{\rm aS/cS}/c$ and the SRS hyperpolarizablity reads $\gamma^{\rm (CRS)} =  (6/4) \ \chi^{(3)}(\Omega_v)$. Because the CARS and CSRS processes are driven by two fields, the expression for the averaged scattered power depends on the process as $p_{\rm aS/cS} = \sigma_{\rm eff}^{\rm (CARS/CSRS)} \, I_{\rm p/S}$. 

A set of label-free interactions fall into the category of pump-probe interactions, which are distinguished by the excitation of a non-equilibrium condition in the system by a first (pump) pulse. That non-equilibrium condition evolves with time and produces a time-varying change in the optical properties of the system that is probed by a second pulse (probe) that arrives at a later time. The excitation by the pump pulse produces a perturbation in the effective linear susceptibility, $\delta \varepsilon_{\rm eff}(t)$ for $t>0$, where we denote $t=0$ as the arrival time of the pump pulse. The time-dependence of the susceptibility perturbation produces spectral scattering that slightly modifies the detected signals. We will neglect the spectral scattering effects, but a recent review of coherent Raman scattering analyses this scenario in detail. \cite{bartels2021low} 

The non-equilibrium condition may be established by the rearrangement of population among electronic or vibrational energy levels. Following the perturbation of the system, the kinetics of the relaxation of the excited state dictates perturbations to the optical properties of the system that can be detected by a time-delayed probe pulse. While these subsequent dynamics can be quite complicated, the effect on the probe pulse can be modeled by a complex-valued $\delta \varepsilon_{\rm eff}(t)$, and the details of the description depend on the detailed spectroscopy interrogated by the probe pulse which can be tuned in wavelength to vary the interaction dynamics. 

In the case of optical absorption induced by a pump pulse, the pump pulse moves population from the ground to an excited electronic state by a change in number density $\delta N$. This population transfer admits several optical spectroscopy perturbations for a time-delayed probe pulse. Details of the particular spectroscopic interactions depend on the center wavelength of the probe pulse. Absorption of the probe pulse can be reduced through ground state depletion or increased through excited state absorption, processes referred to as transient absorption (TA). Alternatively, as some probe wavelengths, a population inversion can be established, leading to stimulated emission (SE) that amplifies the probe pulse. \cite{ye2009nonlinear, min2011coherent, fischer2016invited, wong2021way} The change in susceptibility following pump pulse excitation is given by $\delta \varepsilon_{\rm eff}^{\rm (TA/SA)} = \Delta \alpha^{(1)} \,  \delta N$, for the change in the single-molecule polarizability between the ground and excited states at the probe wavelength, $\Delta \alpha^{(1)}$. In general, $\Delta \alpha^{(1)} = \Delta \alpha_r^{(1)} + i \, \Delta \alpha_i^{(1)} $ is complex valued. When dominated by the imaginary component, $\Delta \alpha_i^{(1)} $, this process is called TA for positive values and called SE for negative values. When the real component, $\Delta \alpha_r^{(1)} $, dominates, the population change is detected through a phase modulation. In all cases, the susceptibility perturbation drives a change in the scattering from the molecule. \cite{wong2021way} The signal change to the probe pulse is causes either gain or loss in the probe field. For the signal model in \reqn{n}, TA and SE make use of the scattering cross-sections from the volume given by $\sigma_{\rm eff}^{\rm (TA/SE)} = (k_{\rm pr}^4 \, V^2/6 \, \pi) \, \left|\delta \varepsilon_{\rm eff}^{\rm (TA/SA)} \right|^2 = \sigma_{\rm eff}^{\rm (TA/SE)}$. The wavenumber of the probe pulse $k_{\rm pr} = \omega_{\rm pr}/c$, and $B^{\rm (TA/SE)}=1$, $g^{\rm (TA/SE)}=1$, and $h^{\rm (TA/SE)}=1$.

As noted above, the change in population $\delta N$ is a perturbation away from thermal equilibrium. This change in population density for an probe pulse with a fluence well below saturation can be computed for a square pulse with peak intensity $I_0$ and pulse duration $\tau_p$, have  $\delta N = \rho \, N$. parameter $\rho = \tau_p \, \lambda \, I_0 \, \sigma_a^{(1)} / h \, c$, where $h$ is Planck’s constant. In the case of two-photon absorption, $\rho = \tau_p \, \lambda \, I_0^2 \, \sigma_{\rm 2PA}^{(1)} / h \, c$, where $\sigma_{\rm 2PA}^{(1)} $ is the 2PA cross-section for a single molecule in units of m$^2$/W. This excited state excitation will relax back to the ground state to reach thermal equilibrium. While this energy decay is often described by a single exponential decay with a excited state lifetime, $\tau_e$, on the order of a few picoseconds for non-fluorescent chromophores, the decay dynamics vary across molecular systems and can be extremely complicated. In addition, this relaxation will lead to thermalization of energy deposited in the molecule with the surrounding environment that can also be used for label-free imaging through the PT detection as described below. 

Pump-probe interactions are also used for vibrational spectroscopic measurements. While vibrational effects, and thus vibrational spectroscopy, can be extracted from dynamics on the excited state of molecules, in label-free microscopy vibrational spectroscopy is usually probed on the ground state through SRS. The processes of SRS is driven by pulses overlapped in time and produces two processes that occur at the same time as CARS and CSRS scattering. SRS, however, produces both loss at the pump frequency driven by the intensity of the Stokes field, leading to stimulated Raman loss (SRL), and gain at the Stokes frequency, leading to stimulated Raman gain (SRG).

Both SRL an SRG produce an effective instantaneous linear susceptibility change that may be written as $\delta \epsilon_{\rm eff}^{\rm (SRL)} = N \, B^{\rm (SRL)} \, \gamma^{(CRS)}$ and $\delta \epsilon_{\rm eff}^{\rm (SRG)} = N \, B^{\rm (SRG)} \, \gamma^{(CRS)*}$. Here $B^{\rm (SRL/SRG)} = (2/\varepsilon_0 \, c) \, \sqrt{g^{\rm (3)}} \,  I_{\rm aS/ap}$, which depends on the product of the average power of either the pump or Stokes beam. The average SRG and SRL power scattered is $p_{\rm SRG/SRL} = \mp \, \sigma_{\rm eff}^{\rm (SRL/SRG)} \, I_{\rm p/S}$. Because the vibrationally resonant contribution to the third-order susceptibility is purely imaginary at peak excitation, $\chi^{(3)} \sim i  (\, N \, \varepsilon_0 / 12 \, m \, \Omega_v \, \Gamma_v) \, \alpha’^2 \sim i \, 3 \times 10^{-20} $ m$^2$ /V$^2$, $\delta \epsilon_{\rm eff}^{\rm (SRL/SRG)}$ is purely imaginary and thus behaves analogously to TA for SRL and SE for SRG. \cite{rigneault2018tutorial} Here, $m$ is the reduced mass of the vibrational mode and $\Gamma_v $ is the linewidth of the vibrational mode resonance. SRL and SRG modify the pump and the Stokes beams through absorption and scattering, and where $\alpha^{\rm (SRL)} = \gamma^{\rm (SRS)}$ and $\alpha^{\rm (SRG)} = \gamma^{\rm (SRS)*}$. Finally, we note that  $B^{\rm (SRL/SRG)} = (2/\varepsilon_0 \, c) \, \sqrt{g^{(3)}} \, I_{\rm aS/ap}$.

SRS is usually implemented with laser pulses longer that the decay time of the excited vibrational coherence. When a pulse duration is shorted so that the $\tau_p \, \Omega_v \ll 1$, the vibrations are driven impulsively by drawing pump and Stokes frequencies from within the bandwidth of a single pump pulse. This limit is referred to as impulsive stimulated Raman scattering (ISRS). \cite{bartels2021low} A vibrational coherence is prepared in the molecule following interaction with a short pump pulse. This vibrational coherence in the impulsive excitation limit produces an oscillating polarizability the leads to an optical susceptibility perturbation of $\delta \varepsilon^{\rm (ISRS)}_{\rm eff} = N \,  B^{\rm (ISRS)} \, \mathrm{Im} \{ \gamma^{(CRS)} \}$. In the impulsive case, we have $B^{\rm (ISRS)} = (3/\varepsilon_0 \, c) \, (\Gamma_v/\nu_r)) \,  I_{\rm a,pu}$, and $I_{\rm a,pu} = p_{\rm pu}/A_{\rm pu}$ denotes the average intensity of pump pulse and $\Gamma_v$ is the decay rate of the vibrational coherence. As with TA/SE, a time-delayed probe pulse interacts with the susceptibility perturbation to produce linear scattering from a spherical particle with polarizability $\alpha^{\rm (ISRS)} = V \, \delta \varepsilon^{\rm (ISRS)}$, and with the usual scattering cross-section, so that we use $p_{\rm pr}$ for \reqn{n}.

The final pump-probe interaction that we consider is the PT effect in which a local change in temperature leads to a change in the local optical susceptibility, $\delta \varepsilon_{\rm eff}^{\rm (PT)} = \Delta T \, (\partial \, \varepsilon/\partial \, T)$ that modifies linear scattering for a probe pulse identical to the cases of any optical excitation process, including absorption or inelastic scattering. The induced perturbation depends on thermal transport because the susceptibility change depends on the change in temperature, $\Delta T$, that is driven by heating from energy deposited into the system and the heat capacity of the medium. As we are considering a system where the target but un-labeled molecules are confined in the volume of a sphere with radius $a \ll \lambda$, we can model the optical response with point heating. On timescales short compared to the thermal transport time, we can estimate the local temperature rise from the energy deposited per excitation $\mathcal{E}_{\rm ex}  = \hbar \,  \Omega_{\rm ex}$, where $\Omega_{\rm ex}$ is the energy gap between ground and excited states. These states can be electronic \cite{tokeshi2001determination, boyer2002photothermal, berciaud2006photothermal, gaiduk2010room} or vibrational. \cite{bai2021bond, zhang2019bond} The total change in energy for $M_{\rm ex}$ states is given as $\Delta Q = \mathcal{E}_{\rm ex}  \, M_{\rm ex} \, \eta_{\rm nr}$. The temperature rise from the heating by the thermal relaxation to the surroundings of the energy deposited in the excitations within a volume, $V$, is $\Delta T = \Delta Q / C_v \,  V$, where $C_v$ is the heat capacity per unit volume of the solvent surrounding the absorber. Nonradiative relaxation of this energy leads to local heating, producing the change in optical susceptibility that is exploited by PT detection. For the case of linear optical absorption, on average the efficiency of nonradiative relaxation $\eta_{\rm nr} = k_{\rm nr}/( k_{\rm nr} + k_{\rm r}) = 1 - \Phi_f$, which is the complement to the quantum fluorescent yield of a molecule, determines the fraction of the energy absorbed by the molecule that contributes to heating. Thus, non-fluorescing chromophores are the best candidates for PT detection. The nonradiative and radiative relaxation rates are $k_{\rm nr}$ and $k_{\rm r}$, respectively. The excited state lifetime of a molecule, given by $\tau_e = (k_{\rm nr} + k_{\rm r})^{-1}$ and which is on the order of several picoseconds for non-fluorescing molecules or several nanoseconds for fluorescent molecules, sets the times scales for the population kinetics following excitation. 

The thermal timescales in a biological imaging scenario are dominated by conductive heat transport. Conductive thermal transport is modelled by the diffusion equation, which gives a diffusion radius $L_{\rm th} = \sqrt{4 \, D \, t}$ in an infinite thermal medium, where $t$ is the time after the point heating has occurred. The thermal transport of the heat away from the absorbers depends on the thermal conductivity, $\kappa = D \, C_v$, and the diffusion coefficient. As we are considering the detection of sub-wavelength particles, we can establish a thermal time scale for diffusion over a wavelength, set by $t_{\rm th} = \lambda^2/4 \, D$. Using a typical value for the diffusion coefficient, $D \sim 10^{-6}$ m$^2$/s, and $\lambda = 10^{-6}$ m, we obtain $t_{\rm th} = 25 \, \mu$s. This timescale is much larger than the pulse spacing in a typical modelocked oscillator, $t_{\rm th} \gg \nu_r^{-1}$, so we may treat the heating and detection with the average beam powers. To eliminate the effects of stray background absorption and scattering, the heating beam is modulated with frequency $\nu_{\rm mod}$ and the thermal transport length associated with this modulation frequency gives a radius of $r_{\rm th} = \sqrt{D/ \pi \, \nu_{\rm mod} }$, which we consider for defining an effective volume of the heated region that induces scattering on the probe pulse. 

On a time interval $\Delta t$ short compared to $t_{\rm th}$, where $\Delta t \ll t_{\rm th}$, there is little time for heat to diffuse as $r_{\rm th}$ will be much less that $\lambda$. However, given such short times, we can estimate the heading that is produced by  $M_p = \Delta t \, \nu_r$ pulses in the time interval for a pulsed source with a repetition frequency of $\nu_r$. The heating per pulse, $\Delta Q$, will accumulate to a total temperature rise of $\Delta T = \Delta Q \, M_p/ C_v \, V $. For optical absorption well below saturation, the mean number of molecules excited by a square heating pulse of length $\tau_h$ and peak intensity $I_{\rm 0h}$ is given by $M_{\rm ex} = \tau_h \, \sigma_a^{(1)} \, I_{\rm 0h} \, N \, V / \mathcal{E}_{\rm ex}$. Putting this together, we obtain a susceptibility perturbation of $\delta \varepsilon_{\rm eff}^{\rm PT} = N \, \delta \varepsilon_i^{(1)} \, B_{\rm PT}$. Here, the imaginary component of the single-molecule susceptibility is $ \delta \varepsilon_i^{(1)} = \sigma_a^{(1)} \, k_0^{-1}$ and the photothermal factor is $B_{\rm PT} = (2/3) \, e^{-2} \, (A_a/A_h) \, (\Delta t_h \, p_h/ \lambda \, \kappa) \, (\partial \, \varepsilon/\partial \, T)$. The heating depends on the ratio of the sphere, $A_a = \pi \, a^2$, to the heating beam cross-section area, $A_h$, the ratio of the average power of the heating beam, $p_h$, to the product $\lambda \, \kappa$. Heating is converted into a change in optical susceptibility through $(\partial \, \varepsilon/\partial \, T)$. The thermal properties of the solvent that scale the PT susceptibility perturbation are $\kappa^{-1} \, (\partial \, \varepsilon/\partial \, T)$. Using numbers from the literature, \cite{gaiduk2010detection} we find that this figure of merit is $\sim 6.6 \times$ larger for glycerol than it is for water, which is why PT imaging experiments use glycerol as a solvent when possible. \cite{zhu2023stimulated}

\renewcommand{\arraystretch}{3} 
\begin{table*}[h!]
    \centering
    \begin{tabular}{|r|c|c|c|}
    \hline
    & $\delta\varepsilon_{\rm eff}$ &  $B$  & $\delta\varepsilon_{\rm{eff},0}$ \\ 
    \hline\hline
    1PA & $\dfrac{M\,\sigma_{a}^{(1)}}{k_{0}\,V}$ & $-$ & \makecell{Chromophores:\\ VIS/UV: $\sim 10^{-4}-10^{-3}$\\mid-IR: $\sim 10^{-5}-10^{-4}$\\ near-IR: $\sim 10^{-6}$}\\[0.25cm]
    \hline
    SHG & $\dfrac{1}{2}B\,\chi_{nr}^{(2)}$ & $\sqrt{\dfrac{2}{\varepsilon_{0}\,c}}h^{(2)}\sqrt{\dfrac{p_{i}}{A_{b}}}$  &\makecell{Collagen: \\ $3.80\times 10^{-4}/mW$}\\[0.25cm]
    \hline
    THG & $\dfrac{1}{4}B\,\chi_{nr}^{(3)}$ & $\dfrac{2}{n_{b}^{2}\,\varepsilon_{0}\,c}h^{(3)}\dfrac{p_{i}}{A_{b}}$ &\makecell{Neat Acetonitrile: \\ $1.86\times 10^{-4}/mW$}\\[0.25cm]
    \hline
    SR & $\dfrac{\sqrt{6\,\pi\,M\,\sigma_{R}^{(1)}}}{V\,k_{s}^{2}}$ & $-$ & \makecell{Non-Res: $\sim 10^{-11}-10^{-10}$\\ Res: $\sim 10^{-8}-10^{-6}$}\\[0.25cm]
    \hline
    CARS/CSRS & $\dfrac{6}{4}B\,\left\vert\chi^{(3)}(\Omega_{\nu})\right\vert$ & $\dfrac{2}{n_{b}^{3/2}\,\varepsilon_{0}\,c}h^{(3)}\dfrac{\sqrt{p_{p}\,p_{S}}}{A_{b}}$ & \makecell{Neat Acetonitrile:\\ $7.06\times 10^{-5}/mW$ }\\[0.25cm]
    \hline
    SRL/SRG & $\dfrac{3}{2}B\,\chi^{(3)/(3)*}_{vr}(\Omega_{\nu})$& $\dfrac{2}{n_{b}^{3/2}\,\varepsilon_{0}\,c}h^{(3)}\dfrac{p_{p}/p_{S}}{A_{b}}$ & \makecell{Neat Acetonitrile:\\ $i7.06\times 10^{-5}/mW$ }\\[0.25cm]
    \hline
    ISRS & $-\dfrac{3}{2}B\,\rm{Im}\left\{\chi^{(3)}_{vr}(\Omega_{\nu})\right\}\dfrac{\Gamma_{\nu}}{\nu_{r}}$ & $\dfrac{2}{n_{b}\,\varepsilon_{0}\,c}h^{(3)}\dfrac{p_{p}}{A_{b}}$ & \makecell{Neat Acetonitrile:\\ $1.18\times 10^{-4}/mW$ }\\[0.25cm]
    \hline
    TA/SE & $\dfrac{\sigma_{a}^{(1)}\,\lambda}{\nu_{r}\,h\,c}\dfrac{p_{i}}{A_{b}}\chi_{vr}^{(1)}$ & $-$ & $-$\\[0.25cm]
    \hline
    PT & $\dfrac{M}{V}\dfrac{\sigma^{(1)}_{a}}{k_{0}}B$ & $\dfrac{2}{3}e^{-2}\dfrac{\pi\,a^{2}}{A_{b}}\dfrac{\Delta t_{h}\,p_{h}}{\lambda\,\kappa}\dfrac{\delta\varepsilon}{\delta T}$ & $-$\\[0.25cm]
    \hline
    \end{tabular}
    
    \caption{Summary of the susceptibility perturbation for various label-free imaging modalities discussed above: 1PA = one photon absorption, SHG = second harmonic generation, THG = third harmonic generation, SR = spontaneous Raman, CARS = coherent anti-Stokes Raman scattering, CSRS = coherent Stokes Raman scattering, SRL = stimulated Raman loss, SRG = stimulated Raman gain, ISRS = impulsive stimulated Raman scattering, TA = transient absorption, SE = stimulated emission, and PT = photothermal. Typical numbers are calculated based on pulsed laser in fs-regime for harmonic generations as well as ISRS and ps-regime for other types of pump-probe Raman methods as well as TA/SE and PT (acts as a CW source).}
    \label{tab:susceptibility_comp}
\end{table*}

\section{Single pixel detection of different label-free imaging methods}

It is evident upon inspection of \reqn{n} that many experimental modalities are admitted by this expression. Having established the general model for the estimation of either a real- or imaginary-valued optical susceptibility perturbation, we will now study the relative performance of methods and comment on the detection sensitivity of various optical methods under some arbitrary susceptibility. We begin with the case of direct signal detection without the aid of interferometric enhancement that is enabled by mixing with a coherent reference field. With this baseline established, we will examine the benefit of reducing noise power via the elimination of the incident field in detection, alternatively known as dark-field imaging.

While experimentally challenging due to the potential weakness of the signal, dark-field imaging oftentimes utilizes a detection angle different from the incident angle. The normalized Fisher information, for both cases of purely real- and imaginary-valued susceptibilities, is expressed as 
\begin{equation}
\label{eq:JnDarkField}
    \tilde{J}_{\rm dark}=\eta_{c}\,\frac{k_{0}^{4}\,V^{2}}{3\,\pi\,A_{b}}.
\end{equation}
It is independent of the object susceptibility and increases as the object volume increases or as the incident beam area decreases. For a typical experimental setup using the dark-field imaging scheme, the Fisher information suggests that $\sim 10^{2}-10^{3}$ photons shall be collected to detect a small susceptibility perturbation as discussed in Sec. \ref{sec:eff_sus}.

Bright-field imaging, by contrast, is a more commonly employed technique in optical experiments. However, the presence of the incident field, whether or not it interferes with the object field, leads to an elevation in the noise level. Increased noise decreases the Fisher information and this is more evident in the real-valued case thanks to the absence of interference between the incident and scattered fields. The corresponding normalized Fisher information expression is given by $\eta_{c}\,k_{0}^{4}\,V^{2}/(\overline{F}_{N,\rm{inc}}^{(\rm im)}+3\,\pi\,A_{b})$, where $\overline{F}_{N,\rm{inc}}^{(\rm im)}$ denotes the scaled incident intensity normalized by the scattered intensity. Here, the information about the object susceptibility is preserved in the scattered field, similar to dark-field imaging, while the noise level experiences an increase. For the imaginary-valued case, although  the interference term contributes additional information about the object susceptibility, the concurrent rise in background results in a higher noise level and consequently, a reduction in the Fisher information. 

\begin{figure}[b!]
    \centering
    \includegraphics[width=\textwidth]{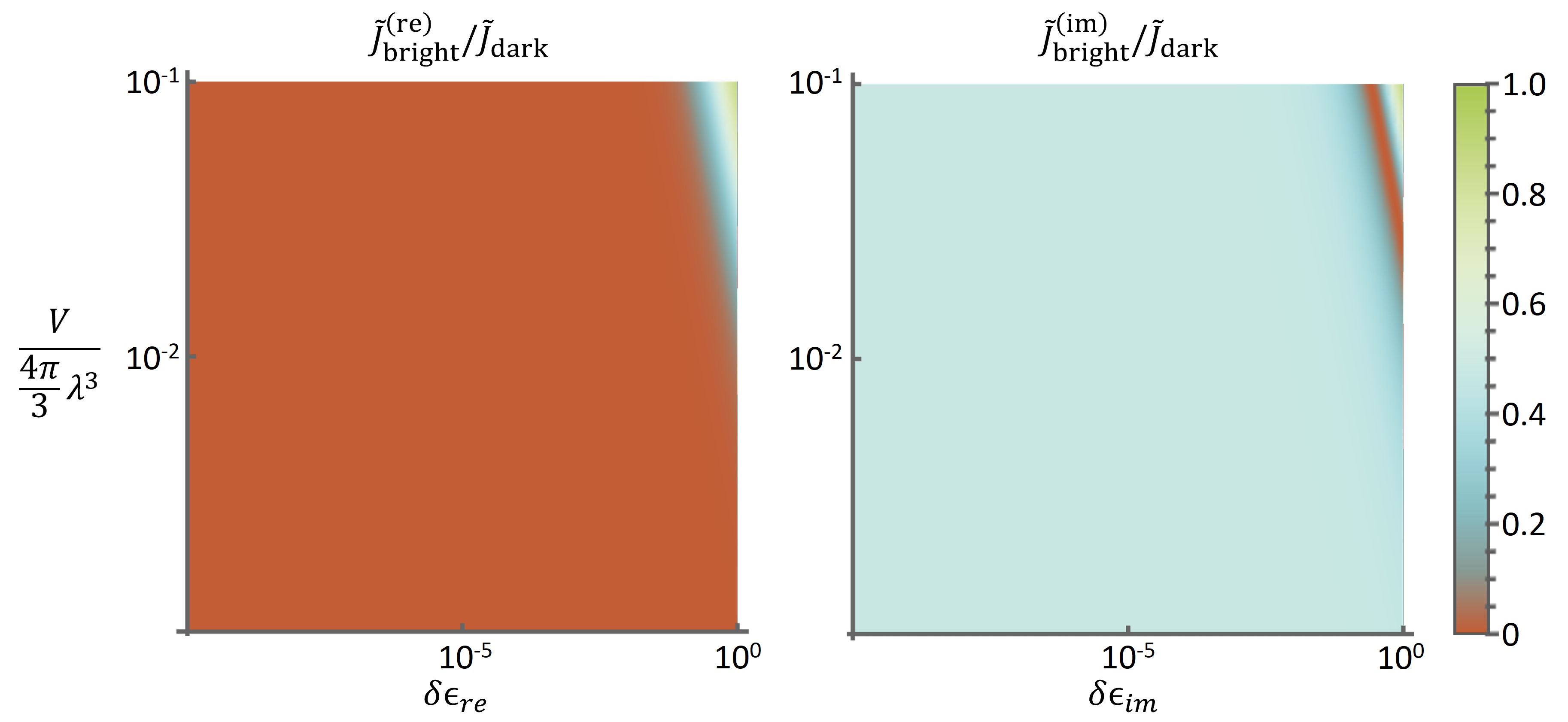}
    \caption{The ratio between the normalized Fisher information in bright-field imaging and in dark-field imaging is plotted as a function of the object susceptibility and volume. The particle optical susceptibility is arbitrarily chosen to be between $10^{-10}$ and $10^{-2}$. Particles are assumed to be suspended in  a sphere of diameter that follows the Rayleigh scattering regime, on the order of less than $\frac{1}{15}$ of the wavelength of the incident. }
    \label{fig:bright_noref}
\end{figure}

However, with an increase in the strength of the object field---whether due to a higher object susceptibility and/or a larger volume---the Fisher information approaches that in dark-field imaging, as illustrated in Fig. \ref{fig:bright_noref}. This phenomenon occurs because the signal from the object field becomes more dominant. For more realistic values of optical susceptibility and volume, the normalized Fisher information in bright-field imaging never exceeds approximately $\ll 1\%$ of that in dark-field imaging for purely real susceptibility perturbation and $\sim 50\%$ for purely imaginary susceptibility perturbation.

\begin{figure}
    \centering
    \includegraphics[width=\textwidth]{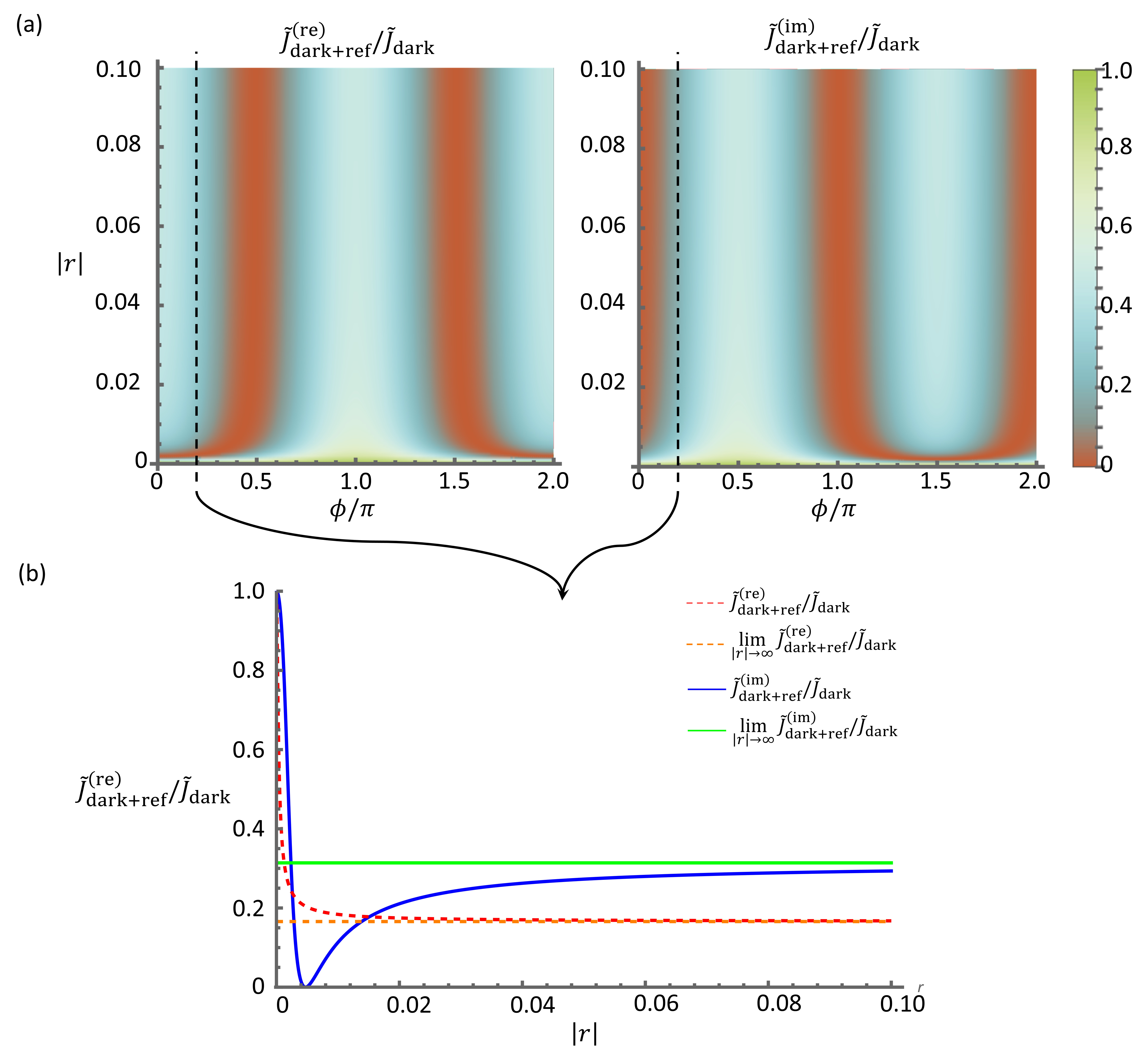}
    \caption{The ratio between the normalized Fisher information in dark-field imaging with reference and in dark-field imaging. (a) Plotted as a function of the relative amplitude and phase of the reference. (b) Plotted as a function of the relative amplitude of the reference to demonstrate its asymptotic behavior as compared to the corresponding approximated limit expression. }
    \label{fig:dark_ref}
\end{figure}

\begin{figure}[b]
    \centering
    \includegraphics[width=\textwidth]{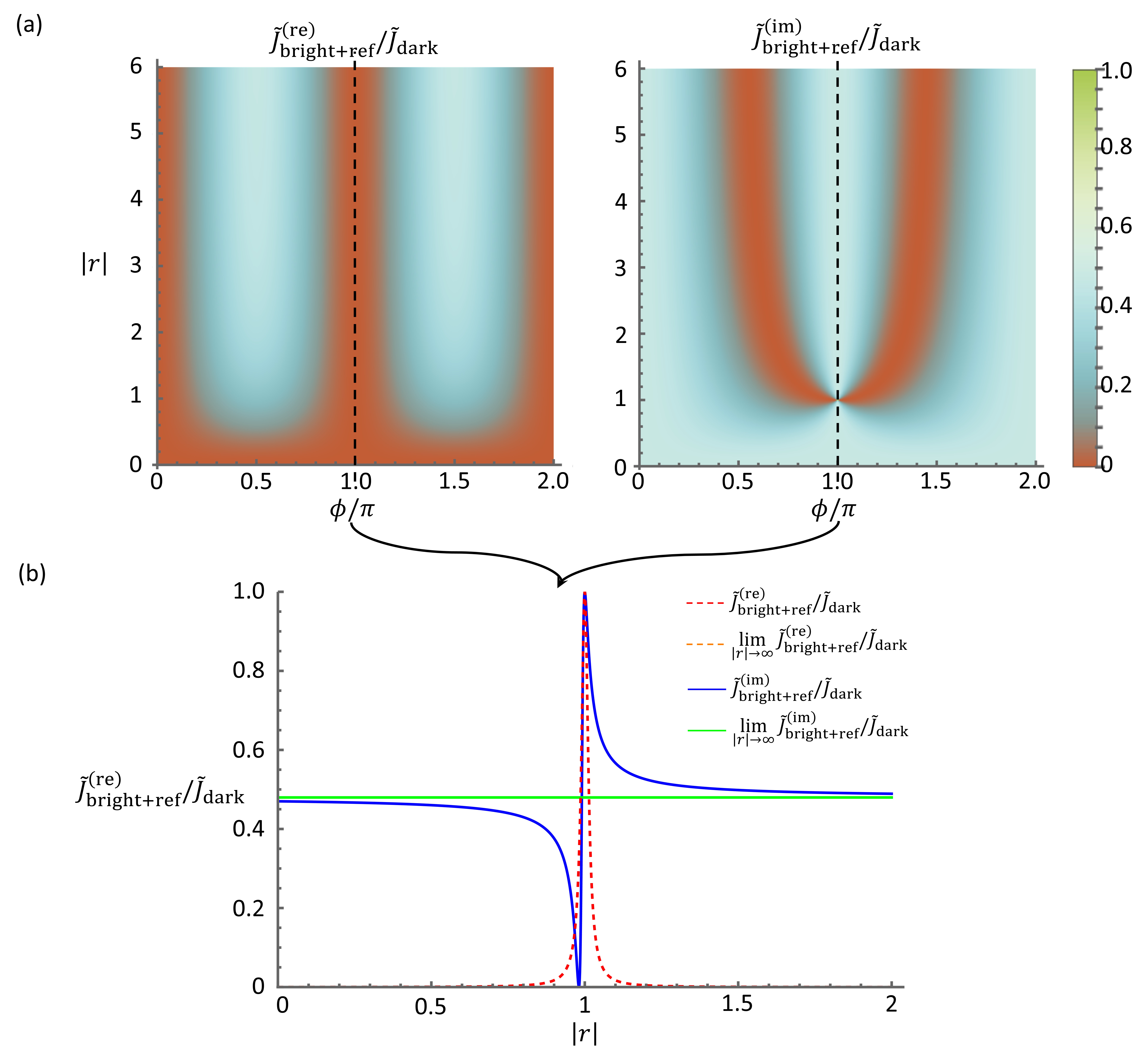}
    \caption{The ratio between the normalized Fisher information in bright-field imaging with reference and in dark-field imaging. (a) Plotted as a function of the relative amplitude and phase of the reference. (b) Plotted as a function of the relative amplitude of the reference to demonstrate its asymptotic behavior as compared to the corresponding approximated limit expression.}
    \label{fig:bright_ref}
\end{figure}

Unlike conventional direct imaging approaches, the exploration of interferometric detection methods in optical experiments opens new avenues for precision and sensitivity. Subtle variations in the optical path length can be precisely detected, enabling the measurement of quantities such as phase differences with exceptional accuracy. To validate its potential enhancement on the estimation of the object susceptibility, we examine the Fisher information in both dark- and bright-field imaging with a coherent reference beam. The ratio of the Fisher information between each detection scheme with reference and the dark-field without reference is plotted in Figs. \ref{fig:dark_ref}(a) and \ref{fig:bright_ref}(a) accordingly.

The relative phase of the reference proves significant, influencing constructive or destructive interference with the object field. The interference results in a boost or diminishment in information about the optical susceptibility perturbation without drastically changing the overall measurement and consequently, the noise level. Therefore, the Fisher information analyses allow for determining the optimal reference beam intensity and phase relative to the incident.

Given the demonstrated enhancement in sensitivity achieved by suppressing the noise level, it is only natural to question whether dark-field imaging with reference offers any advantages over its bright-field counterpart. Interestingly, as long as the reference is not fully eliminated, the noise level experiences similar increases in both detection schemes, while the information pertaining to the susceptibility perturbation remains constant. This leads to comparable reductions in the Fisher information. However, as the reference strength increases, the interference term gains dominance, leading to the asymptotic behavior of the normalized Fisher information converging to
\begin{equation}
\label{eq:ref_Asp_re}
    \lim_{|r|\rightarrow\infty} \tilde{J}_{\rm ref}^{(\rm re)} = \frac{k_{0}^{2}\,V^{2}\,f^{(j)}\sin{(\phi)}}{A_{b}^{2}},
\end{equation}
for the real-valued susceptibility case and for the imaginary-valued susceptibility case,
\begin{equation}
\label{eq:ref_Asp_im}
    \lim_{|r|\rightarrow\infty} \tilde{J}_{\rm ref}^{(\rm im)} = \frac{k_{0}^{2}\,V^{2}\,f^{(j)}\cos{(\phi)}}{A_{b}^{2}}
\end{equation}
in both schemes. The demonstration of such asymptotic behavior is illustrated by plotting the Fisher information against the relative amplitude and phase of the reference in Figs. \ref{fig:dark_ref}(b) and \ref{fig:bright_ref}(b). Notably, the convergence rate is influenced by the object susceptibility or volume, with slower convergence observed for larger values of these parameters.  Nonetheless, the ratio between the Fisher information in either detection scheme with reference and dark-imaging converges rapidly to $\sim50\%$ at optimal relative reference phases. 

More incidentally, introducing a reference beam in the bright-field imaging not only increases the noise level, leading to a loss of information, but also reveals the potential for destructive interference between the incident and reference fields, effectively suppressing the background and thereby reducing the noise level. Particularly intriguing is the scenario where the reference completely cancels out the incident, resulting in an equivalent to dark-field imaging, and maximizing the Fisher information, as illustrated in Fig. \ref{fig:bright_ref}(b).

In summary, our findings align with the initial assessment derived from the Fisher information expression in Sec. \ref{sec:prob_crlb}. We have demonstrated that lower noise power leads to reduced relative variability, consequently enhancing precision in estimating object susceptibility. Specifically, the maximum Fisher information is attained in dark-field imaging without reference and equivalently, in bright-field imaging with reference when the reference cancels out the incident. In cases where a reference beam is necessary for signal detection facilitation, opting for a relatively large beam intensity is advisable to achieve improved estimation precision. However, it's worth noting that the best achievable precision in optical susceptibility is $\sim\sqrt{2}\times$ worse than that in dark-field imaging without reference.

\section{Conclusions}

We have presented a comprehensive examination of signal detection methods in label-free imaging, culminating in the development of a universal signal model for measuring the optical susceptibility of sub-wavelength particles applicable across various imaging modalities. By leveraging Fisher information analyses, we have explored the sensitivity of each modality, assuming limiting cases in optical susceptibility, purely real or imaginary, and noise, shot-noise-limited.

Our Fisher information analyses suggest that dark-field imaging is the most sensitive, yielding the best precision in estimating the particle optical susceptibility, due to the lack of noise contributions from incident or reference sources. This finding is prevalent with the presence of other noises while the signal sits above the additive noise floor. Moreover, we propose an approach to achieve dark-field-like imaging using bright field imaging with a reference arm, where the reference effectively cancels out the incident light in both amplitude and phase. The advantage with the reference is the boost in signal intensity to mitigate the weak scattered light from the sample. 

In this calculation, we considered the limiting case of ideal optical detection, where we need only consider shot noise in the detection. Such a model neglects the effects of noise in the detector, such as Johnson noise and dark current, as well as relative intensity noise (RIN) that is present in optical sources. We make these assumptions because we are primarily interested in the limiting case of weak signal detection, and thus low dark-field signal flux. At such low signal levels, RIN becomes negligible, and the dominant noise that we must contend with is dark current noise. Dark current is modeled as a Poisson process and thus is an additive noise contribution in the denominator of the Fisher information. The key issue is that the dark counts must be lower than the dark-field signal flux. Under such conditions, the conclusions that we draw here are widely applicable. 

Building on the legacy left behind by Gabi Popescu, future research endeavors may focus on practical implementations of dark-field-like imaging approach, while further exploration of its applications in diverse scientific and biomedical domains remains crucial.

\bibliography{LabelFreeCRLB}

\end{document}